\documentclass[aps,prl,showpacs,twocolumn,groupedaddress,a4paper]{revtex4}

\usepackage{graphics}

\begin{document}


\title{Simulation of Transitions between ``Pasta'' Phases in Dense Matter
}


\author{Gentaro Watanabe$^{a,b,c}$, Toshiki Maruyama$^{d}$,
  Katsuhiko Sato$^{b,e}$, Kenji Yasuoka$^{f}$ and Toshikazu Ebisuzaki$^{c}$}
\affiliation{
$^{a}$NORDITA, Blegdamsvej 17, DK-2100 Copenhagen \O, Denmark
\\
$^{b}$Department of Physics, University of Tokyo,
Tokyo 113-0033, Japan
\\
$^{c}$The Institute of Chemical and Physical Research (RIKEN),
Saitama 351-0198, Japan
\\
$^{d}$ASRC, Japan Atomic Energy Research Institute, Tokai,
Ibaraki 319-1195, Japan
\\
$^{e}$Research Center for the Early Universe, University of Tokyo,
Tokyo 113-0033, Japan
\\
$^{f}$Department of Mechanical Engineering, Keio University,
Yokohama, 223-8522, Japan}


\date{\today}

\begin{abstract}
Calculations of equilibrium properties of dense matter predict that at
subnuclear densities
nuclei can be rodlike or slablike.
To investigate whether transitions between phases with non-spherical nuclei 
can occur during the collapse of a star,
we perform quantum molecular dynamic simulations of the
compression of dense matter.  We have succeeded in simulating the 
transitions between rodlike and slablike nuclei and between slablike 
nuclei and cylindrical bubbles.
Our results strongly suggest that non-spherical nuclei can be formed
in the inner cores of collapsing stars.
\end{abstract}


\maketitle


In ordinary matter, atomic nuclei are roughly spherical
because, in the liquid drop picture of the nucleus,
effects of the nuclear surface tension are greater
than those of the Coulomb forces.
When the density of matter approaches that of atomic nuclei,
calculations predict that, in equilibrium state,
the nuclei will adopt different shapes, such as 
cylinders and slabs, etc.
These phases with non-spherical nuclei are often referred 
to as ``pasta'' phases \cite{rpw,hashimoto}.

In the initial stage of the supernova explosions,
matter in the collapsing iron cores experiences an adiabatic compression,
which leads to an increase of the density in the central region
from $\sim 10^{9}$ g cm$^{-3}$ to around the normal nuclear density
$\rho_{0}=0.165$ fm$^{-3}$ just before the star rebounds;
the temperature there reaches the order of 1 MeV.
The pasta phases are thus expected to be formed in the inner cores
during the collapse of stars.
However, such a speculation is based on phase diagrams of the equilibrium state
(e.g., Refs.\ \cite{lassaut,qmd_hot} for finite temperatures)
or static and perturbative calculations \cite{review,iida}.
It is still unclear whether or not the pasta phases can be formed
and the transitions between them can be realized during the collapse,
which lasts less than a second.
Because of the drastic changes of nuclear shape that occur under
non-equilibrium conditions,
this problem is more difficult than the realization of the pasta phases
by cooling at constant density, as demonstrated in Ref.\ \cite{qmd},
and an {\it ab-initio} approach is called for.

In the present Letter, we solve the problem
about the transitions between pasta phases
using a dynamical framework for nucleon many-body systems
called the quantum molecular dynamics (QMD) \cite{aichelin}.
QMD is a suitable approach to describe thermal fluctuations and 
is efficient enough to treat large systems
consisting of several nuclei.
Furthermore, at the relevant
temperatures of several MeV, shell effects, which cannot be
described by QMD, are less important because they washed out
by thermal fluctuations.

The pasta phases
have recently begun to attract the attention of researchers
(see, e.g., Ref.\ \cite{burrows} and references therein).
The mechanism of the collapse-driven supernova explosion
has been a central mystery in astrophysics for almost half a century.
Previous studies suggest that
the revival of the shock wave
by neutrino heating is a crucial process.
As has been pointed out in Refs.\ \cite{gentaro2,qmd}
and elaborated in Refs.\ \cite{horowitz2,prep},
the existence of the pasta phases
instead of uniform nuclear matter increases
the neutrino opacity of matter in the inner core significantly
\cite{note_opacity}
due to the neutrino coherent scattering by nuclei
\cite{freedman,sato1};
this affects the total energy transferred to the shocked matter.
Thus the pasta phases could play an important role in the future study
of supernova explosions.

In the present study, we use a nuclear force given by a QMD Hamiltonian
with medium-EOS parameter set in Ref.\ \cite{maruyama}.
This Hamiltonian contains the momentum dependent ``Pauli potential'',
which reproduces the effects of
the Pauli principle phenomenologically.
Parameters in the other terms of the Hamiltonian are determined to reproduce
the saturation properties and the properties of
finite nuclei in the ground state,
especially of heavier ones relevant to the present study \cite{maruyama}.
It is also confirmed that a QMD Hamiltonian close to the present model
provides a good description of nuclear reactions including the low energy
region (several MeV per nucleon) \cite{niita}, which would be important
for the present case.

Using the above QMD Hamiltonian,
we perform simulations of symmetric nuclear matter with 16384 nucleons
in a cubic box with periodic boundary condition
(see Ref.\ \cite{prep} for other cases).
The system contains equal numbers of protons (and neutrons)
with spin up and spin down.
The relativistic degenerate electrons which ensure charge neutrality are
treated as a uniform background because, at subnuclear densities,
the effect of the electron screening is small \cite{review,screening}.
The Coulomb energy,
taking account of the Gaussian charge distribution of the proton wave packets,
is calculated by the Ewald method.
The temperature $T$ is measured by the effective kinetic temperature
for momentum-dependent potentials,
which is consistent with the temperature in the Boltzmann statistics
\cite{qmd}.
The QMD equations of motion
are integrated by the fourth-order Gear predictor-corrector method
with a multiple time step algorithm.
Integration time steps $\Delta t$ are adaptive
in the range of $\Delta t < 0.1$ -- $0.2$ fm$/c$.

As the initial condition, we use samples of the columnar phase
and of the laminar phase of 16384-nucleon system at 
$T\simeq1$ MeV obtained in Ref.\ \cite{qmd_hot}.
In preparing them,
we first combine eight replicated 2048-nucleon samples in the ground state
($T\simeq0$ MeV;
nucleon number density
$\rho=0.225\rho_0$ for the phase with rodlike nuclei
and $0.4\rho_0$ for the case of slablike nuclei) into a 16384-nucleon sample,
and then put random noise on the positions and the momenta of nucleons
up to $0.1$ fm and $1$ MeV$/c$, respectively.
We equilibrate the sample at $T=1$ MeV for $\sim$ 4000 -- 5000 fm$/c$
using the Nos\'e-Hoover thermostat
for momentum-dependent potentials \cite{qmd_hot}.
We further relax the sample for $\sim$ 5000 fm$/c$ without the thermostat.

Starting from the above sample,
we simulate the adiabatic compression.
In the case starting from the phase with rodlike nuclei [slablike nuclei],
the density is increased by 2$\times 10^{-4} \rho_0$
[1$\times10^{-4} \rho_0$] every 100 steps
by changing the box size
(the particle positions are rescaled at the same time).
The average rate of change for the density is
$\simeq$1.3$\times 10^{-5} \rho_0/$(fm$/c$)
[$\simeq$7.1$\times 10^{-6} \rho_0/$(fm$/c$)];
this rate ensures the adiabaticity of the simulated compression process
with respect to the change of nuclear structure \cite{adiabat}.
Finally, we relax the compressed sample
at $\rho=$0.405 $\rho_0$ [0.490 $\rho_0$].
These final densities are those
of the phase with slablike nuclei [cylindrical bubbles]
in the equilibrium phase diagram at $T \simeq 1$ MeV \cite{qmd_hot}.

The resulting
time evolution of the nucleon distribution
is shown in Figs.\ 1 and 2.
Starting from the phase with rodlike nuclei 
[Fig.\ 1-(1); $\rho=0.225 \rho_0$ (volume fraction 
of nuclear matter region $u=0.20$ -- 0.22 \cite{volume}) at time $t=0$],
the phase with one-dimensional layered lattice
of slablike nuclei is formed 
[Fig.\ 1-(9); $\rho=0.405 \rho_0$ ($u=0.41$ -- 0.45) at $t=17720$ fm$/c$].
During the compression, the temperature increases gradually
up to $\simeq 1.35$ MeV in the final state.
We note that, in the process of the compression,
the phase with rodlike nuclei persists as a metastable state,
and moreover, until nuclei begin to touch and fuse 
they are not elongated along the plane of
the final slabs [see Fig.\ 1-(2)].
This shows that the transition from the phase with rodlike nuclei
to the slablike nuclei is not triggered by the fission instability.
This result is consistent with a previous study,
which shows the stability of the rodlike nuclei
against a small quadrupolar deformation
of the cross section \cite{iida}.
When the internuclear spacing becomes small enough 
and once some pair of neighboring rodlike nuclei 
touch due to thermal fluctuations,
they fuse [see the lower two nuclei in the middle column in Figs.\ 1-(3)
and 1-(4); $u=0.27$ -- 0.30 at $t=6050$ fm$/c$].
Like a chain reaction,
such connected pairs of rodlike nuclei further touch and fuse with 
neighboring nuclei in the same lattice plane
[see Figs.\ 1-(5) and 1-(6)].
Each fusion process in the chain reaction proceeds
on a time scale of order $10^2$ fm$/c$,
which is much shorter than the time scale of
the density change \cite{adiabat}.

\begin{figure}
\resizebox{8.5cm}{!}
{\includegraphics{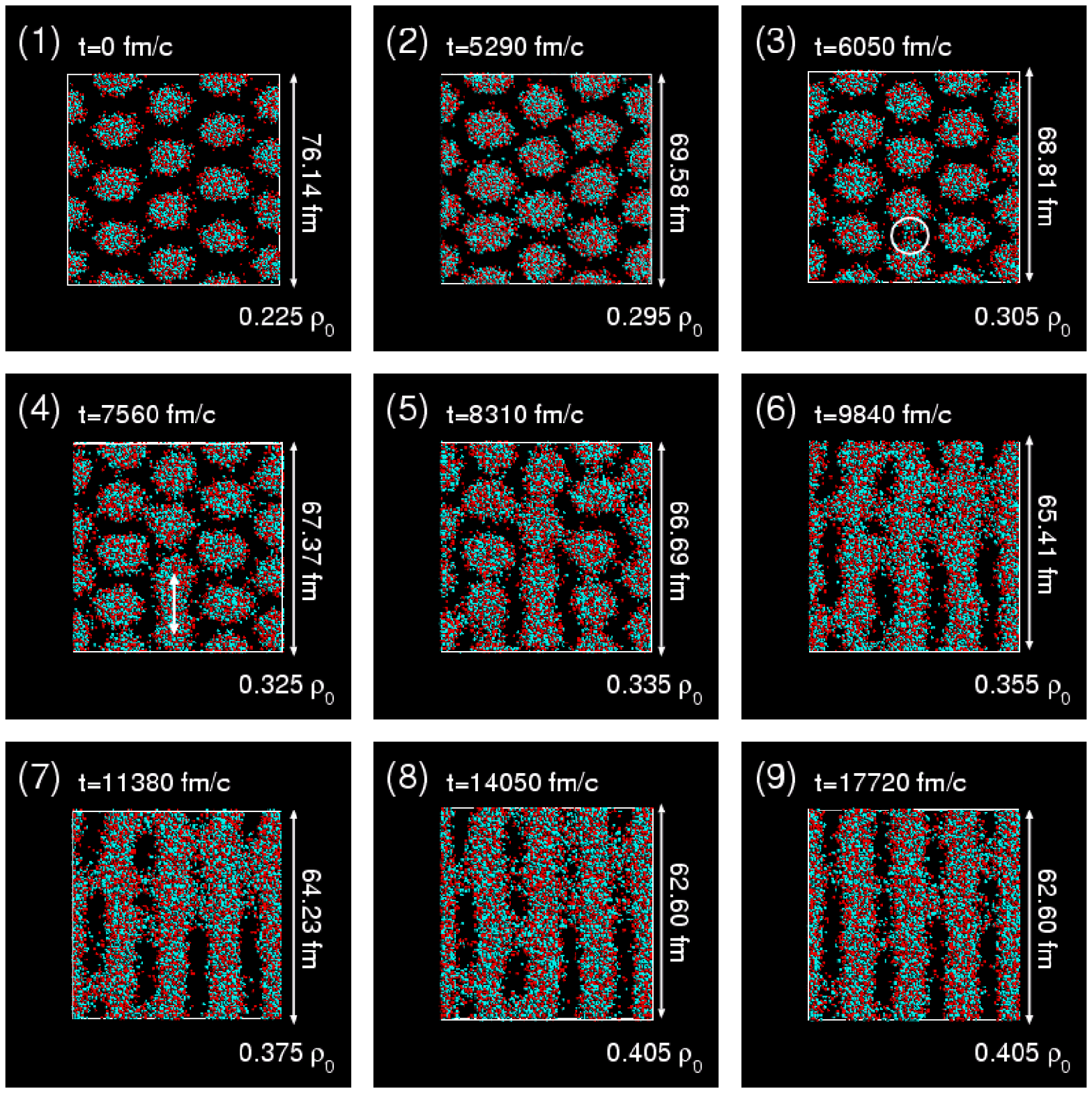}}
\caption{(Color) Snapshots of the transition process from
the phase with rodlike nuclei to the phase with slablike nuclei
(the whole simulation box is shown).
The red particles show protons and the green ones neutrons.
After neighboring nuclei touch as shown by the circle in Fig.\ 1-(3),
the ``compound nucleus'' elongates along the arrow in Fig.\ 1-(4).
The box size is rescaled to be equal in this figure.
}
\vspace{0.5cm}
\resizebox{8.5cm}{!}
{\includegraphics{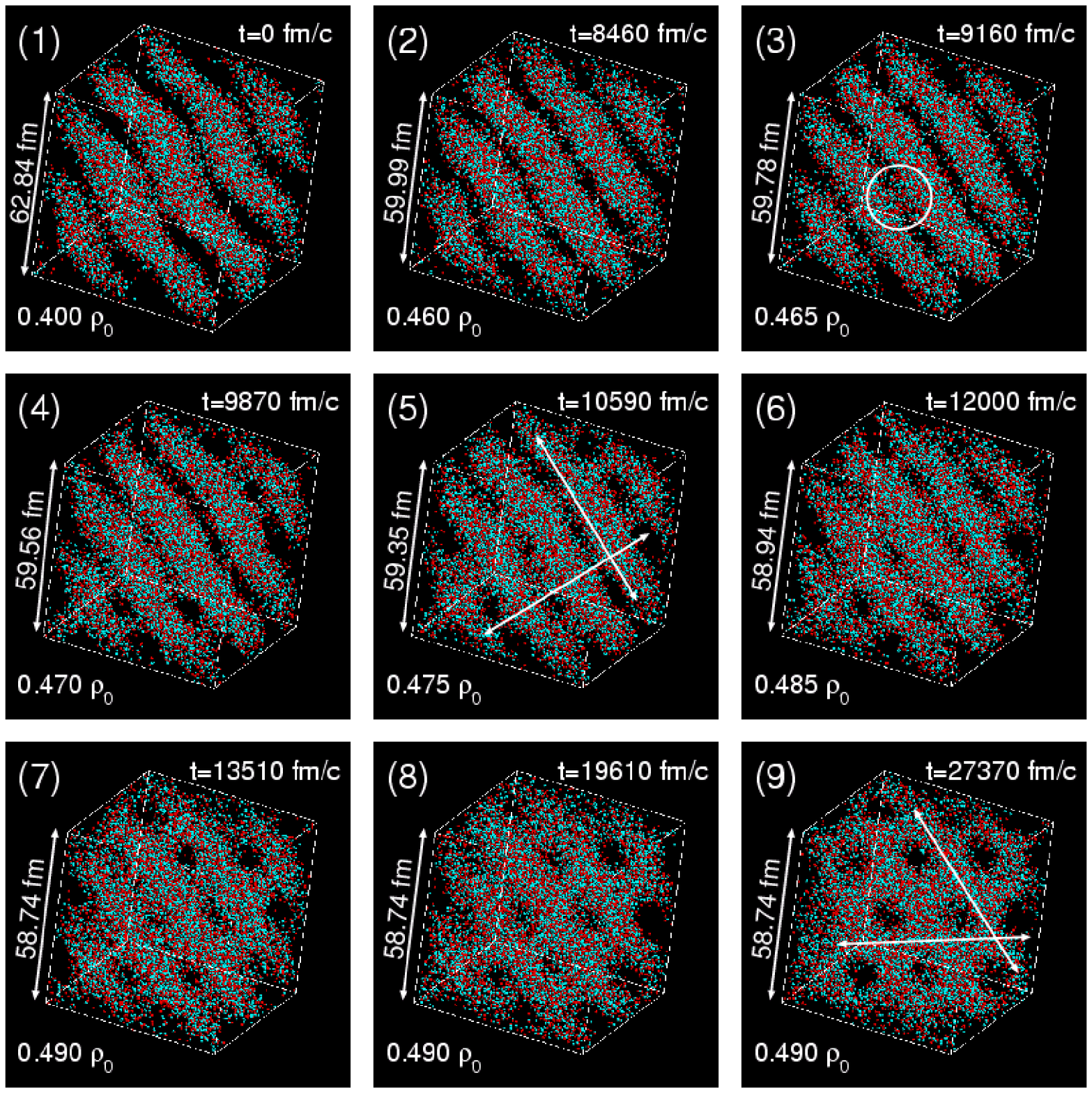}}
\caption{(Color) The same as Fig.\ 1 for the transition from
the phase with slablike nuclei to the phase with cylindrical holes
(the box size is not rescaled in this figure).
After the slablike nuclei begin to touch [see the circle in Fig.\ 2-(3)],
the bridges first crosses them almost orthogonally as shown by the arrows
in Fig.\ 2-(5). Then the cylindrical holes are formed and 
they relax into a triangular lattice, as shown by the arrows in Fig.\ 2-(9).
}
\end{figure}

The transition from the phase with slablike nuclei
to the phase with cylindrical holes is shown in Fig.\ 2
($u=0.42$ -- 0.45 and 0.55 -- 0.59 at $t=0$ and 27370 fm$/c$).
When the internuclear spacing decreases enough,
neighboring slablike nuclei touch due to the thermal fluctuation
as in the above case.
Once nuclei begin to touch ($u=0.49$ -- 0.52 at $t=8460$ fm$/c$),
bridges between the slabs are formed at many places
on a time scale (of order 100 fm$/c$)
much shorter than that of the compression [cf. Figs.\ 2-(3) and 2-(4)].
After that the bridges cross the slabs
nearly orthogonally for a while,
which makes hollow regions on a square lattice rather than the final triangular one.
Nucleons in the slabs continuously flow into the bridges,
which become wider and merge together to form cylindrical holes.
Afterwards, the connecting regions
consisting of the merged bridges move gradually,
and the cylindrical holes relax to form a triangular lattice.
The final temperature in this case
is $\simeq 1.3$ MeV.

Let us now investigate 
the detailed time evolution of the nuclear structure.
The integral mean curvature and the Euler characteristic
(see, e.g., Ref.\ \cite{minkowski})
are powerful tools for this purpose.
Suppose there is a set of regions $R$,
where the density is higher than
a threshold density $\rho_{\rm th}$. 
The integral mean curvature and the Euler characteristic
for the surface of this region $\partial R$
are defined as surface integrals of
the mean curvature $H$ and
the Gaussian curvature $G$, respectively;
i.e., $\int_{\partial R} H dA$ and
$\chi \equiv \frac{1}{2 \pi} \int_{\partial R} G dA$,
where $dA$ is the area element of the surface of $R$.
The topological quantity
$
  \chi = \mbox{(number of isolated regions)}
  - \mbox{(number of tunnels)} + \mbox{(number of cavities)}\
  \label{euler}
$
\cite{note_euler}.
We calculate these quantities using the density field of nucleons
on $128^3$ grid points
(see Ref.\ \cite{qmd} for detailed procedures).

\begin{figure}
\rotatebox{270}{
\resizebox{!}{8.8cm}
{\includegraphics{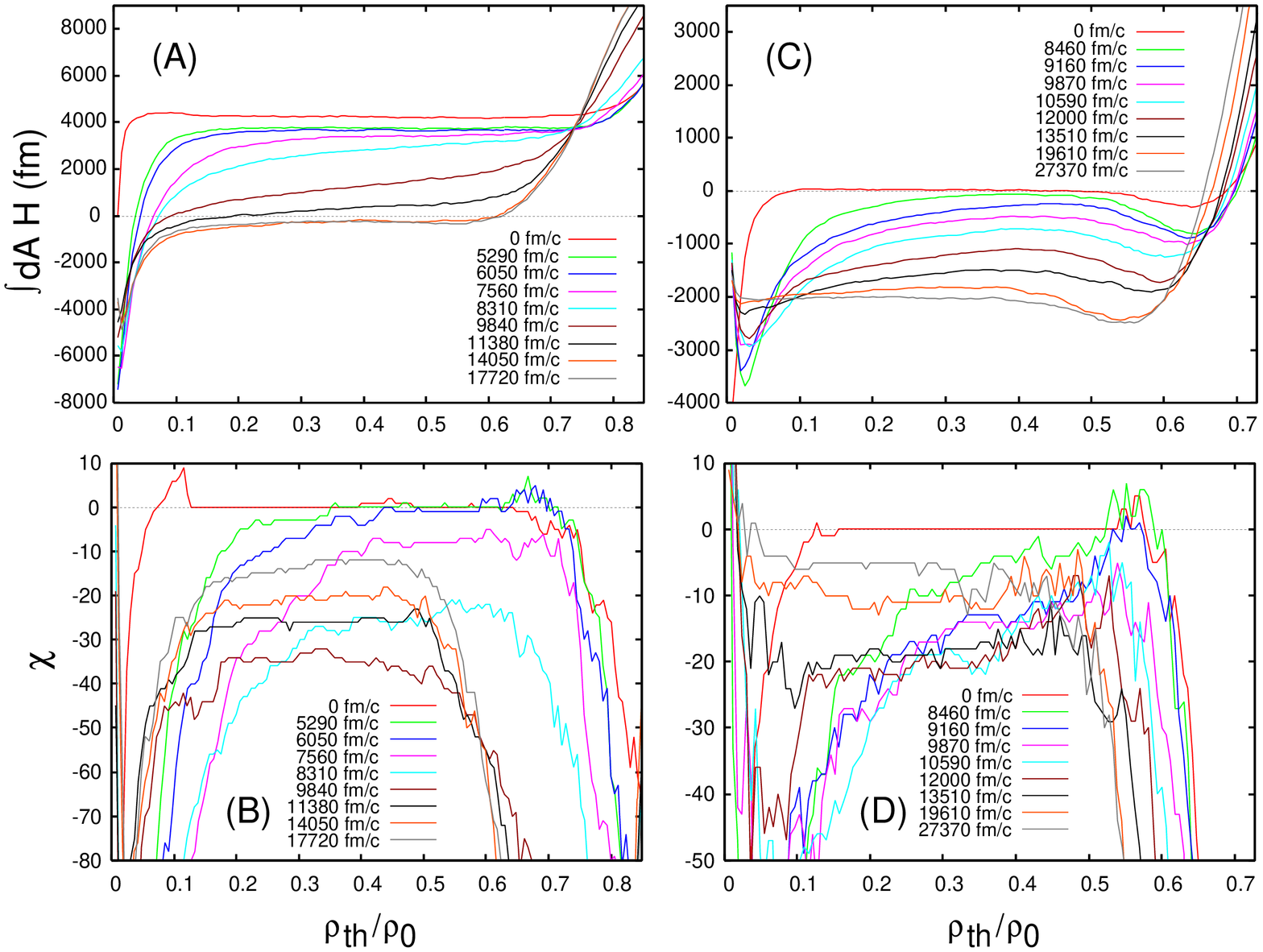}}}
\caption{(Color) The quantities $\int_{\partial R} H dA$ and $\chi$
as functions of the threshold density 
$\rho_{\rm th}$ calculated for the nucleon density fields of Fig.\ 1
[(A) and (B)] and of Fig.\ 2 [(C) and (D)].
}
\rotatebox{270}{
\resizebox{!}{8.8cm}
{\includegraphics{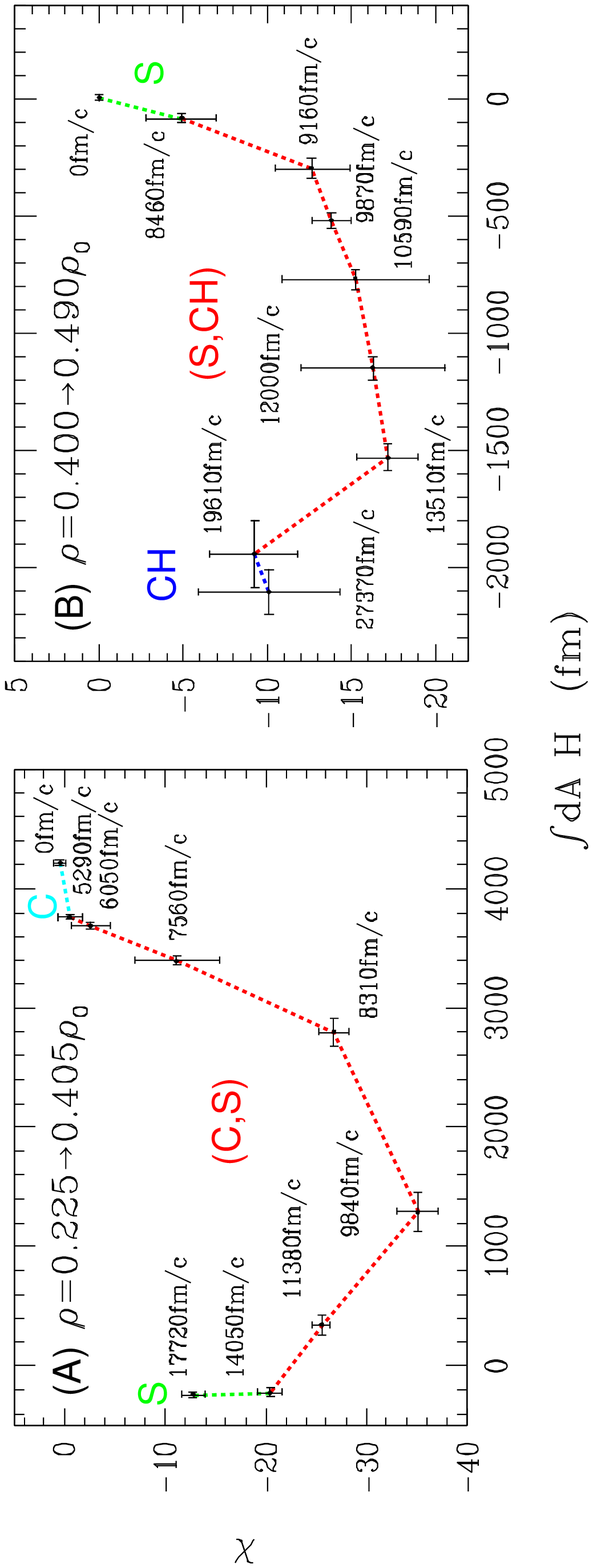}}}
\caption{(Color) Time evolution of $\int_{\partial R} H dA$ and $\chi$
during the simulations.
The data points and the error bars show, respectively, the mean values
and the standard deviations in the range of $\rho_{\rm th}=0.3$ -- $0.5 \rho_0$,
which includes the values corresponding to the nuclear surface assumed in
\cite{volume}.
We set the averaging range relatively wide because the nuclear surface
cannot always be characterized by a single value of $\rho_{\rm th}$
as mentioned in \cite{volume}.
The panel (A) is for the transition from cylindrical [C]
to slablike nuclei [S]
and the panel (B) for the transition from slablike nuclei
to cylindrical holes [CH].
Transient states are shown as (C,S) and (S,CH) for each transition.
}
\end{figure}

In Figs.\ 3-(A) and 3-(B), the quantities $\int_{\partial R} H dA$ and $\chi$
calculated for each nucleon distribution in Fig.\ 1
are shown as functions of $\rho_{\rm th}$.
In this case, the initial condition is the phase with rodlike nuclei,
whose structure is well characterized by the plateau of
$\int_{\partial R} H dA \simeq 4000$ fm
[see the red curve in Fig.\ 3-(A)].
The slablike nuclei in the final state,
on the other hand, are characterized by
$\int_{\partial R} H dA \simeq 0$ fm,
corresponding to the plateau of the gray curve in Fig.\ 3-(A).

The behavior of $\chi$ clearly shows that the rodlike nuclei begin to touch
between $t=5290$ and 6050 fm$/c$, when the plateau value of $\chi$
starts to deviate from zero, characterizing the rodlike nuclei,
to negative values, characterizing the multiply connected structures 
such as sponges.
We note that before the nuclei touch,
the change in $\int_{\partial R} H dA$
is small except for lower values of $\rho_{\rm th} \lesssim 0.1 \rho_0$.
This reflects the facts that the phase with rodlike nuclei
persists as a metastable state and that the transition is not induced by
the fission instability \cite{note_fission}.
Also the behavior of $\chi$ should be noted;
as can be seen from Fig.\ 4-(A),
it becomes negative between the phase with rodlike nuclei
($\chi\simeq 0$ and $\int_{\partial R}HdA >0$)
and the phase with slablike nuclei
($\chi\simeq 0$ and $\int_{\partial R}HdA \simeq 0$).
This implies that the transition proceeds
through a transient state with ``spongelike'' structure.
The state which gives the smallest $\chi$ at $t\simeq 9840$ fm$/c$ 
corresponds to the moment
when all of the rodlike nuclei are connected to others
by small bridges;
after that, the connected nuclear rods relax into slablike nuclei,
i.e., the bridges in the slablike structures merge to form the nuclear slabs
and those across the slabs disappear.
The whole transition process can be divided into
the ``connecting stage'' and the ``relaxation stage''
before and after this moment;
the former starts when the nuclei begin to touch and it
takes $\simeq 3000$ -- 4000 fm$/c$ and the latter takes
more than 8000 fm$/c$.

The same quantities
are shown for the transition
from the phase with slablike nuclei to the phase with cylindrical bubbles
in Figs.\ 3-(C), 3-(D), and 4-(B).
The initial and the final structures are characterized by the plateau values
of $\int_{\partial R} H dA \simeq 0$ and $\simeq -2000$ fm, respectively.
From Figs.\ 3-(D), and 4-(B),
we see that the slablike nuclei
begin to touch at $t \lesssim 8460$ fm$/c$ and
the connection of the slablike nuclei by the small bridges are
completed at $t \lesssim 12000$ fm$/c$ corresponding to the state
with the lowest $\chi$.
In this case, the connecting stage lasts for
$\simeq 3000$ -- $4000$ fm$/c$
and the relaxation stage for more than $15000$ fm$/c$.
In the latter period, the bridges merge to form cylindrical holes
shown by the increase of $\chi$ toward zero, and, simultaneously,
their positions relax into a triangular lattice as mentioned before.

In conclusion, we have succeeded in simulating the dynamical process of
two types of transitions between pasta phases at subnuclear densities.
Our calculations support the idea that transitions between pasta phases
can occur during stellar collapse.
The particular transitions we have examined are triggered by
the thermal fluctuation, not by the fission instability.
They consist of the connecting stage and the relaxation stage.
The total time of the connecting stage is $3000$ -- $4000$ fm$/c$
in our simulations, which could be shortened by the artificial
compression. However, we can conclude that
the connecting stage would be complete
in a time scale of order $10^3$ fm$/c$
taking account of the facts that
each connecting process observed in this stage proceeds much faster than
the compression and that the time scale of ordinary nuclear fission
is about $1000$ fm$/c$.
The relaxation stage takes about $10000$ fm$/c$ or more.
A remaining challenge is to investigate the transition
from the bcc lattice of spherical nuclei to the triangular lattice of
rodlike nuclei \cite{prep}.
If this process is confirmed,
the existence of the pasta phases in supernova cores
will be almost established.

\begin{acknowledgments}
G. W. is grateful to C. J. Pethick and K. Iida
for valuable discussions and comments.
G. W. also appreciates
M. Shimizu, L. M. Jensen, P. Urkedal, T. Iitaka and T. Shimobaba.
T. M. thanks S. Chiba for discussions and kind suggestions.
This work was supported in part
by the Nishina Memorial Foundation,
by the Ministry of
Education, Culture, Sports, Science and Technology
through Research Grant No. S14102004, No. 14079202 and No. 14-7939,
and by RIKEN through Research Grant No. J130026.
\end{acknowledgments}


\begin{thebibliography}{99}
%
\bibitem{rpw} D. G. Ravenhall, C. J. Pethick and J. R. Wilson,
  Phys.\ Rev.\ Lett. {\bf 50}, 2066 (1983).
%
\bibitem{hashimoto} M. Hashimoto, H. Seki and M. Yamada,
  Prog.\ Theor.\ Phys. {\bf 71}, 320 (1984).
%
\bibitem{lassaut} M. Lassaut {\it et al.},
  Astron.\ Astrophys. {\bf 183}, L3 (1987).
%
\bibitem{qmd_hot} G. Watanabe {\it et al.},
  Phys.\ Rev.\ C {\bf 69}, 055805 (2004).
%
\bibitem{review} C. J. Pethick and D. G. Ravenhall,
  Annu.\ Rev.\ Nucl.\ Part.\ Sci. {\bf 45}, 429 (1995).
%
\bibitem{iida} K. Iida, G. Watanabe and K. Sato,
  Prog.\ Theor.\ Phys. {\bf 106}, 551 (2001).
%
\bibitem{qmd} G. Watanabe {\it et al.},
  Phys.\ Rev.\ C {\bf 66}, 012801(R) (2002); {\it ibid.},
  {\bf 68}, 035806 (2003).
%
\bibitem{aichelin} J. Aichelin,
  Phys.\ Rep. {\bf 202}, 233 (1991).
%
\bibitem{burrows} A. Burrows, S. Reddy and T. A. Thompson,
  Nucl.\ Phys. {\bf A}, in press (astro-ph/0404432).
%
\bibitem{gentaro2} G. Watanabe, K. Iida and K. Sato,
  Nucl.\ Phys. {\bf A687}, 512 (2001);
  Erratum, Nucl.\ Phys. {\bf A726}, 357 (2003).
%
\bibitem{horowitz2} C. J. Horowitz, M. A. P\'erez-Garc\'ia, and J. Piekarewicz,
  Phys.\ Rev.\ C {\bf 69}, 045804 (2004).
%
\bibitem{prep} G. Watanabe {\it et al.}, in preparation.
%
\bibitem{note_opacity} The cross section for the neutrino coherent scattering
is approximated to be proportional to the static structure factor
of neutrons, whose peak height for the columnar and laminar phases
($x=0.3$, $T=1$ MeV) is $O(10^2)$ \cite{prep}.
%
\bibitem{freedman} D. Z. Freedman,
  Phys.\ Rev.\ D {\bf 9}, 1389 (1974).
%
\bibitem{sato1} K. Sato,
  Prog.\ Theor.\ Phys. {\bf 53}, 595 (1975).
%
\bibitem{maruyama} T. Maruyama {\it et al.},
  Phys.\ Rev.\ C {\bf 57}, 655 (1998).
%
\bibitem{niita} K. Niita, JAERI-conf. {\bf 96-009}, 22 (1996) (in Japanese).
%
\bibitem{screening} G. Watanabe and K. Iida,
  Phys.\ Rev.\ C {\bf 68}, 045801 (2003).
%
\bibitem{adiabat} According to the typical value of the density difference
between each pasta phase, $\sim 0.1\rho_0$,
we increase the density to the value corresponding to the next pasta phase
taking the order of $10^4$ fm$/c$, which is much longer than
the typical time scale of the nuclear fission, $\sim 1000$ fm$/c$.
Therefore, if a dynamical phenomenon of nuclei whose time scale is less than
the order of $10^4$ fm$/c$ is observed during the compression,
its dynamics is determined by the intrinsic physical properties
of the system not by the density change applied externally,
whose time scale is much shorter than that of the actual stellar collapse.
%
\bibitem{volume} We here assume that the nuclear surface corresponds to
the isodensity surface for a threshold density (see later) of
$\rho_{\rm th}=0.45$ -- $0.5 \rho_0$.
We should mention that the nuclear surface cannot be characterized by
a single value of $\rho_{\rm th}$ except for the state before the fusion
and after the system is fully relaxed because the density in the nuclear
matter region in the transient states is quite inhomogeneous.
%
\bibitem{minkowski} K. Michielsen and H. De Raedt,
  Phys.\ Rep.\ {\bf 347}, 461 (2001).
%
\bibitem{note_euler} The value of $\chi$
is positive for the phases with spherical nuclei and spherical bubbles,
and is zero for the other ideal pasta phases, i.e.,
the phases of rodlike nuclei, slablike nuclei, and cylindrical bubbles.
%
\bibitem{note_fission} $\int_{\partial R}H dA$
should decrease before $\chi$ deviates from zero 
if the transition is triggered by the fission instability,
i.e., nuclei deform largely before they touch.
We see, however, only a minor change of 
$\int_{\partial R}HdA$ before $\chi$ decreases.
%
\end{thebibliography}
\end{document}